\documentclass[aip,jcp,reprint,a4paper,floatfix,showkeys,tighten]{revtex4-1}
\usepackage{natbib}
\usepackage[dvips]{graphicx}
\DeclareGraphicsExtensions{.eps}
\usepackage{color}
\usepackage{amsmath}
\usepackage{amssymb}
\usepackage{bm}
\usepackage{dcolumn}
\usepackage{bm}
\newcommand{\un}[1]{\ensuremath{\unskip\,\mathrm{#1}}}
\newcommand{\dd}{\textrm d}
\newcommand{\beq}{\begin{equation}}
\newcommand{\eeq}{\end{equation}}

\begin{document}

\title{The interaction of hybrid nano-particles inserted within surfactant bilayers}

\author{Doru Constantin}
\email{constantin@lps.u-psud.fr}
\affiliation{Laboratoire de Physique des Solides, Universit\'{e}
Paris-Sud, CNRS, UMR 8502, 91405 Orsay, France.}

\date{7 October 2010}

\begin{abstract}
We determine by small-angle X-ray scattering the structure factor of hydrophobic particles inserted within lamellar surfactant phases for various particle concentrations. The data is then analyzed by numerically solving the Ornstein-Zernicke equation, taking into account both the intra- and inter-layer interactions. We find that particles within the same layer repel each other and that the interaction potential (taken as independent of the concentration) has a contact value of $2.2\, k_B T$ and a range of about 10~\AA. If the amplitude is allowed to decrease with increasing concentration, the contact value in the dilute limit is about $5\, k_B T$, for a similar range.
\end{abstract}




\maketitle

\section{Introduction}
\label{intro}

The elucidation of membrane-mediated interaction between inclusions
in the cell membrane (such as integral proteins or membrane-active
antibiotic peptides) is of paramount importance for understanding
their biological activity. Since the underlying problem is the
organization of the membrane --seen as a two-dimensional complex
system-- it has become clear over the last three decades that the
concepts developed in soft matter physics for the understanding of
self-assembled systems are operative in this context and that
`simple' models can yield valuable information.

It is therefore not surprising that sustained theoretical efforts
attempted to provide a detailed description of these complex
systems; they are either continuum-elasticity theories
\cite{Dan:1993,Dan:1994,Aranda:1996,Bohinc:2003} or more detailed
models taking into account the molecular structure of the lipid
bilayer \cite{Marcelja:1976,Sintes:1997,Lague:2000,May:2000}. In the
few cases when the models were validated and refined using
experimental data, these approaches were often successful. For
instance, the lifetime of the gramicidin channel is known to depend
on the thickness \cite{Elliott:1983} and tension \cite{Goulian:1998}
of the membrane; it was shown that a continuum-elasticity model
provides a satisfactory description of this phenomenon
\cite{Huang:1986}.

However, this body of theoretical work has not yet been matched by
the experimental results, the first of which were obtained by
directly measuring the radial distribution function of membrane
inclusions using freeze-fracture electron microscopy (FFEM)
\cite{Lewis:1983,Chen:1973,James:1973,Abney:1987}. These data were
compared to liquid state models and could be described by a
hard-core model with, in some cases, an additional repulsive or
attractive interaction \cite{Pearson:1983,Pearson:1984,Braun:1987}.
However, FFEM has not been extensively employed since, certainly due
to the inherent experimental difficulties; furthermore, the particle
distribution observed in the frozen sample is not necessarily
identical to that at thermal equilibrium.

A step forward was taken by Huang and collaborators, who showed that
the distribution of membrane inclusions within the plane of the
layers can be studied using small-angle X-ray and neutron scattering
\cite{Harroun:1999,Yang:1999}. First of all, these techniques are
perfectly adapted to the typical length scales to be probed. The
measurement is averaged over a large number of particles and over
long times, without perturbing the system; thus, one has access to
the structure factor of the interacting particles. However, in these
studies, at most two peptide-to-lipid concentrations $P/L$ were
investigated for each system and no values were given for the
interaction potential, the data being explained in terms of purely
hard-core interactions.

Building upon this work, we recently studied systems with a varying
density of inclusions. This is indispensable since, even though each
structure factor (taken separately) can be described by a hard-core
model, the apparent radius obtained changes with the concentration,
signalling the presence of an additional interaction. For instance,
we were able to measure the interaction potential of alamethicin
pores in dimyristoyl-phosphatidylcholine (DMPC) bilayers. We showed
that, aside from the expected hard-disk repulsion (with a radius
corresponding to the geometrical radius of the pore) the pore
interaction exhibits a repulsive contribution, with a range of about
$3 \un{nm}$ and a contact value of $2.4 \, k_BT$
\cite{Constantin:2007}. On the other hand, for gramicidin pores in
dilauroyl-phosphatidylcholine (DLPC) bilayers, while the interaction
is still repulsive, the parameter values are quite different, with a
higher contact value and much shorter range. In the latter case, the
interaction was shown to decrease with the pore concentration, in
agreement with the hydrophobic matching model
\cite{Constantin:2009}. These results are in qualitative agreement
with recent theoretical predictions \cite{Lague:2000,Lague:2001} and
can be used as a test for other theoretical and numerical results.

The biological molecules cited above act specifically by insertion
within cell membranes, so their affinity for lipid bilayers is
assured, which is a significant advantage. However, their use as
membrane probes has important shortcomings: their X-ray scattering
contrast is low and the positions of the constitutive atoms are not
always well-defined (membrane proteins can adopt various
conformations as a function of the environment, while the pores
formed by antimicrobial peptides often comprise variable numbers of
monomers). These features contribute to the difficulty of obtaining
high-quality scatering data on membranes with inclusions.

Fortunately, if one is interested in the general physical properties
of membranes seen as two-dimensional complex fluids, rather than in
the behaviour of a specific active molecule, other inclusions can be
used. For instance, we have shown recently that tin oxo-clusters can
be inserted within surfactant bilayers and that their interaction
potential can be determined \cite{Constantin:2008}.

These types of hybrid organic-inorganic particles have the following
advantages:

\begin{itemize}

\item They are perfectly monodisperse (actually iso-molecular) and
``rigid'' (their atomic configuration is well-defined).

\item Their scattering contrast is high (due to the presence of metal
atoms).

\item Their surface properties can be tailored by changing the nature
of the peripheral groups.

\end{itemize}

The purpose of this article is to develop a full analysis of the system, beyond the initial approach used in Ref.~\onlinecite{Constantin:2008}. Two main improvements can be noted:

\begin{itemize}

\item We were able to determine the complete structure
factor, $S(q_r,q_z)$, while the initial study was only concerned
with the equatorial slice, $S(q_r,q_z = 0)$. This required new experimental data, for $q_z \neq 0$.

\item We employed a more elaborate statistical model
(based on the numerical solution of the Ornstein-Zernicke equation
with the Percus-Yevick closure) to relate the structure factor to
the interaction potential; our first approach was based on the
random-phase approximation (RPA).

\end{itemize}

We have also checked the reproducibility of the results (see Appendix \ref{reprod}), by comparing the old and new data, thus verifying both the reliability of the method and the stability of the samples.

\section{Materials and Methods}
\label{matmeth}

We used as inclusions hybrid nanoparticles consisting of a tin oxide core decorated with butyl chains. The complete formula is: \{(BuSn)$_{12}$O$_{14}$(OH)$_6$\}$^{2+}$(4-CH$_3$C$_6$H$_4$SO$_3^-$)$_2$, shortened to BuSn12 in this paper. The synthesis and structural details are given in Ref.~\onlinecite{Eychenne:2000}. The BuSn12 particles were dissolved in ethanol prior to use.

The membranes were composed of dimethyldodecylamine-N-oxide (DDAO), a single-chain zwitterionic surfactant. The DDAO (purchased from Sigma-Aldrich) was first dried in vacuum for 20~h to remove any residual water \cite{Kocherbitov:2006} and then dissolved in isopropanol.

The two stock solutions were then mixed to yield the desired BuSn12/DDAO ratio and the mixtures dried in vacuum, yielding a final mass of about 200~mg for each sample. Varying amounts of water were then added so that the mixtures were in the fluid lamellar $L_{\alpha}$ phase \cite{Kocherbitov:2006}. For DDAO, the molecular weight is 229.40, the density is $0.84 \, \text{g/cm}^3$ and the bilayer thickness is $25 \pm 1 \, \text{\AA}$ \cite{Oradd:1995,Wasterby:1998}, resulting in an area per surfactant molecule $A_{\text{DDAO}} = 37.8 \, \text{\AA}^2$. The molecular weight of the BuSn12 is 2866.7, with a density of $1.93 \, \text{g/cm}^3\,$ \cite{Eychenne:2000}. The (two-dimensional) number density of particles in the plane of the membrane, $n$, is calculated using the data above and neglecting the possible increase in bilayer surface due to the inclusions.

The lamellar phases were then drawn into flat glass capillaries (VitroCom Inc.,
Mt. Lks, NJ), 100~$\mu$m thick and 2~mm wide by aspiration with a syringe and the capillaries were flame-sealed. Good homeotropic alignment (lamellae parallel to the
flat faces of the capillary) was obtained by thermal treatment, using a Mettler~FP52 heating stage. The samples were heated up to the isotropic phase (at $130 {\,}^{\circ} \text{C}$) and then cooled down to the lamellar phase at a rate of $1 {\,}^{\circ} \text{C/min}$.

We studied the samples by small-angle X-ray scattering (SAXS), on the the bending magnet beamline BM02 (D2AM) \cite{Simon:1997} of the European Synchrotron Radiation Facility (ESRF, Grenoble, France). The photon energy was 11~keV and the sample-to- detector distance about 27~cm, with a scattering vector in the range $0.04 < q < 0.9 \,
\text{\AA}^{-1}$. The detector is a charge-coupled device Peltier-cooled camera (SCX90-1300, from
Princeton Instruments Inc., NJ) with a detector size of $1340 \times 1300$ pixels. Data preprocessing (dark current subtraction, flat field correction, radial regrouping and normalization) was performed using the \texttt{bm2img} software developed at the beamline. The scattering geometry is discussed in section \ref{results}.

\section{Model and analysis}
\label{analysis}

Since the particles are identical in shape, the scattering intensity
can be written as the product of a form factor (only depending on
the internal constitution of the particles) and a structure factor,
which describes the interaction between particles \cite{Chaikin95}:
$I(\bm{q})=S(\bm{q})\cdot \left | F(\bm{q}) \right |
^2$, with:

\beq \label{eq:struct} S(q_z,q_r)= \frac{1}{N} \left \langle \left |
\sum_{k=1}^{N-1} \exp \left ( - i \bm{q} \bm{r}_k \right )
\right | ^2 \right \rangle \eeq

\noindent where $N$ is the number of objects and object ``0'' is
taken as the origin of the coordinates.

We expect the form factor of the particles to be dominated by their inorganic core, as the electron density of the butyl chains is similar to that of the dodecyl chains within the bilayers and to that of ethanol. This assumption is confirmed by the intensity at higher scattering vectors ($q > 0.5 \, \text{\AA}^{-1}$), which is well described for all samples by the form factor of a sphere $|Ff(R,q)|^2$, with a radius (used as a free fitting parameter) $R = 4.5 \pm 0.2 \, \text{\AA}$, in good agreement with the average radius of the tin oxide ``cage'' estimated from the crystallographic data.

Dividing the measured intensity by the form factor above yields the structure factor $S(\bm{q})$. Standard error propagation then yields the uncertainty, but the resulting values (corresponding to the statistical error) are much smaller than the discrepancy between the fitting model and the experimental data (the goodness-of-fit function ${\chi}^2$ per data point is much larger than one). Recognizing that this discrepancy is mostly due to systematic errors in data acquisition, background subtraction etc. and also possibly to the inadequacy of the theoretical models, we assign to each data point a fixed uncertainty $\sigma$; in the following, we take for definiteness $\sigma = 0.05$, which is a rough estimate of the difference between model and data, insofar it yields ${\chi}^2$ values of unit order.

Determining the structure factor for systems composed of many
particles interacting via a known potential is one of the
fundamental endeavors of liquid state theory.

Two cases must be considered:

\begin{itemize}

\item The situation when the particles are dissolved in a simple
solvent will act as a ``reference state''; more precisely, it will
allow us to check whether the ``naked'' particles can be described
simply by a hard core repulsion or whether an additional term must
be considered and what its characteristics are.

\item The most interesting situation is of course that of particles
inserted within the surfactant bilayers, which will be the main
result of our study. Due to the anisotropic nature of the matrix,
the analysis must account for the difference between the interaction
between particles contained within the same layer and that from one
layer to the next.

\end{itemize}

\subsection{Structure factor of an isotropic solution}
\label{3Dmodel}

In solution, the interaction between particles is taken as
isotropic, described by a two-body potential depending only on the distance
between their centers: $V_{3D}(\bm{r})=V_{3D}(r)$. Consequently,
the structure factor $S(q)$ is equally isotropic, depending only on
the magnitude of the wave vector $q$.

Using as input the interaction potential, the structure factor was
determined numerically by solving iteratively the Ornstein-Zernicke
equation with the Percus-Yevick closure condition. The algorithm is
based on the method of Lado \cite{Lado:1967,Lado:1968}; see Appendix \ref{model} for the details. The interaction potential is described by a hard core with diameter $2R = 9.4 \un{\AA}$ and an additional ``soft'' component, modelled either as a Gaussian centered at the origin\cite{Constantin:2008}: $V(r) = U_0 \exp \left [ - \frac{1}{2}\left ( \frac{r}{\xi} \right )^2 \right ] \quad r \geq 2R$ or as a decreasing exponential with contact value $U_c$ and decay range $\xi$: $V(r) = U_c \exp \left [ - (r - 2R) / \xi \right ] \quad r \geq 2R$.

\begin{figure}
\centerline{\includegraphics[width=9cm,angle=0]{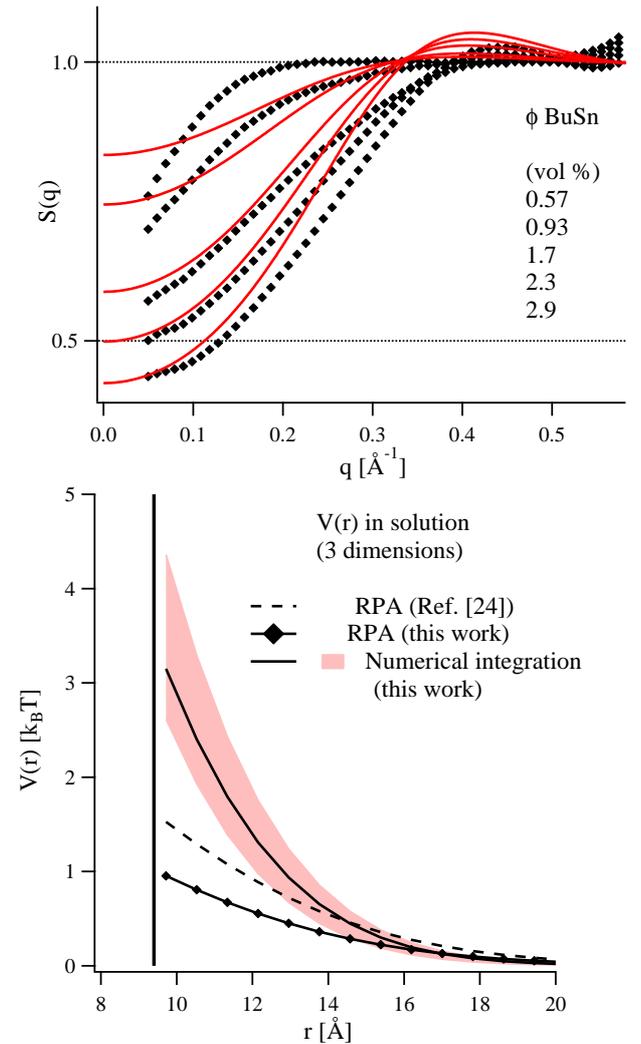}}
\caption{Top: Experimental three-dimensional structure factors for BuSn12 particles in ethanol (symbols) and fits with the Gaussian model discussed in the text (solid lines) for five volume concentrations $\phi$. $S(q)$ decreases with increasing $\phi$. Bottom: Interaction potential obtained from the data above via the numerical procedure discussed in Section~\ref{3Dmodel} (solid line and shaded area for the uncertainty range) and within the RPA approximation (solid line and symbols). For comparison, we also show the RPA results obtained in Ref.~\onlinecite{Constantin:2008} (dashed line).} \label{fig:V_3D}
\end{figure}

\subsection{Structure factor in the lamellar phase}
\label{2Dmodel}

\begin{figure*}[htb]
\centerline{\includegraphics[width=18cm,angle=0]{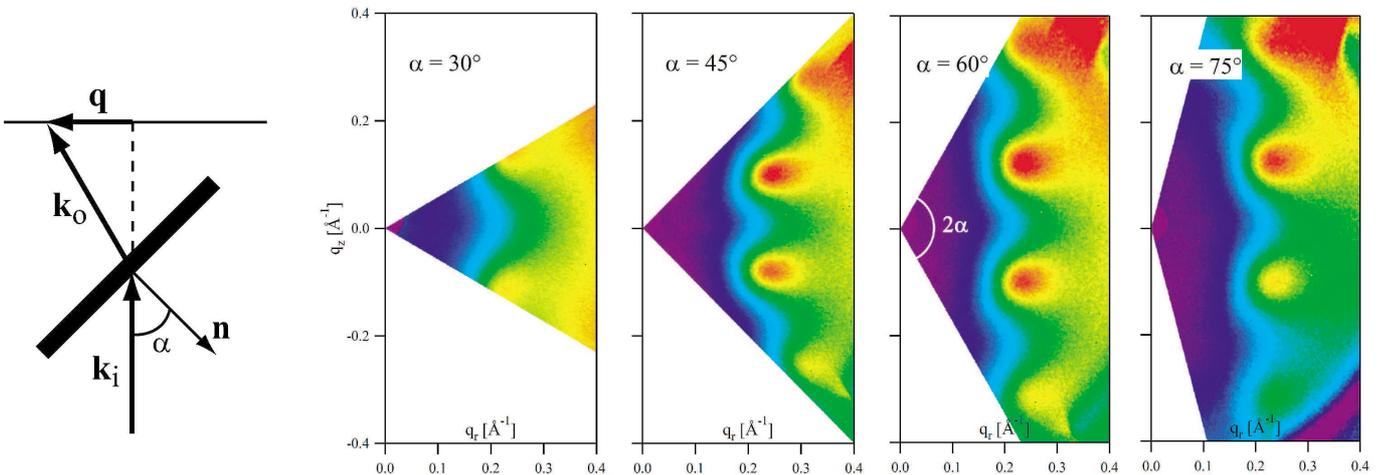}}
\caption{Left: Sketch of the experimental configuration. Right: Re-gridded scattering data for various values of the incident angle $\alpha$ (indicated within the images).} \label{fig:redr}
\end{figure*}

The phase (and hence the structure factor) is now anisotropic. We
assume that there is no in-plane ordering (the particles form a 2D
liquid in the plane of the bilayer), so that $S$ only depends on the
absolute value of the in-plane scattering vector $q_r = \sqrt{q_x^2
+ q_y^2}$ and on the scattering vector along the membrane normal
$q_z$.

For an ideal gas (no interaction) $S=1$, while if the particles
interact only in the plane of the bilayer and not from one bilayer
to the next, $S=S(q_r)$ only. In the general
case, the structure factor can be developed as \cite{Yang:1999}:

\beq \label{eq:develop} S(q_r,q_z) = S_0(q_r)+2 \sum_{m=1}^{\infty}
\cos (q_z d m) S_m (q_r)\eeq

\noindent with each partial structure factor $S_m$ describing the interaction
between particles situated $m$ layers away:

\beq \label{eq:Sm} S_m (q_r) = \delta _{0m} + 2 \pi \rho
\int_{0}^{\infty} r \dd r J_0(r q_r) \left [ g_m (r) -1 \right ]
\eeq

\noindent where $m \geq 0 $ and $g_m (r)$ is the (normalized) probability of
finding a particle at a distance $r$ in bilayer $m$, knowing that a
particle is present at the origin in bilayer 0.

This expansion merely reflects the discrete nature of the stack
along the $z$ direction. It is especially convenient since generally
the interaction does not extend very far along $z$.
Note also that Equation (\ref{eq:develop}) is only valid for the
geometrically perfect case when the distance between neighboring
layers is rigorously $d$. In practice, this distance varies due to
thermal fluctuations and to frozen-in defects, leading to a smearing
of the diffraction pattern at high $q_z$ values. We describe this
effect phenomenologically by a Lorentzian factor. Finally, we use:

\beq \label{eq:develop_final} S(q_r,q_z) = S_0(q_r)+2 \frac{\cos
(q_z d)}{1 + (q_z \sigma)^2} S_1 (q_r)\eeq

\noindent where the disorder parameter $\sigma$ has units of length. For
thermal fluctuations, $\sigma ^2= \langle (z_1 - z_0)^2 \rangle
\simeq \eta (d/\pi)^2$, where $\eta$ is the Caill\'{e} parameter.

As we will see below, the structure factor $S_1 (q_r)$ due to
nearest-bilayer interaction has a localized peak. For
convenience, we describe it by a Gaussian function:

\beq \label{eq:S1} S_1 = A_1 \exp \left [ - \frac{(q -
q_{\un{max}})^2}{2 \, \Delta q ^2} \right ] \eeq

The partial structure factors $S_0(q)$ and $S_1(q)$ are similar to those
describing a two-dimensional binary mixture AB, with the formal
identification $S_0 = S_{AA} = S_{BB}$ and $S_1 = S_{AB}$. As we will see below, these two components are enough to describe the experimental data, so there is no need to go beyond $m=1$ (interaction between adjacent bilayers).

The (experimentally determined) functions $S_0(q)$ and $S_1(q)$ can be described --using well-established results in liquid state theory-- in terms of the interaction potentials $V_0(r)$ (between particles within the same layer) and $V_1(r)$ (between adjacent layers). The derivation of the model and the implementation details are given in Appendix \ref{model}.

\section{Results}
\label{results}

\subsection{Interaction in solution}
\label{solution}

The 3D structure factors of BuSn12 particles in ethanol are shown in Figure~\ref{fig:V_3D} (top) for five different concentrations (lines and symbols). They are fitted with the Gaussian interaction model discussed in Section~\ref{3Dmodel}, yielding $U_0 = 15 \, k_B T$ and $\xi = 5.1 \, \text{\AA}$, such that the potential at contact is $U_c = 3.5 \, k_B T$. The fits are shown as solid lines. The corresponding interaction potential is shown in Figure~\ref{fig:V_3D} (bottom) as solid line. The shaded area represents the uncertainty (see Appendix \ref{subsec:conf} for details). For comparison, we show in the same figure the potential obtained in the RPA approximation for the current data (solid line and symbols) and for that in Ref.~\onlinecite{Constantin:2008} (dashed line). The exponential model yields very similar results, in terms of goodness of fit and in potential shape for parameter values $U_c = 4.5 \, k_B T$ and $\xi = 2.1 \, \text{\AA}$.

\subsection{Interaction between layers}
\label{qz}

In order to quantify the interaction between the layers (along the smectic director $\hat{z}$), one needs access to the complete structure factor $S(q_r,q_z)$. This is achieved by using the experimental configuration described in Figure~\ref{fig:redr}, as first discussed in Ref.~\onlinecite{Yang:1999}: the incoming X-ray beam (with wave vector $\bm{k}_i$) is incident upon the flat capillary at an angle $\alpha$ to its normal $\bm{n}$, which coincides with the smectic director $\hat{z}$. A point on the 2D detector uniquely defines an outgoing wave vector $\bm{k}_o$ for the scattered signal, and thus a scattering vector given by: $\bm{q}=\bm{k}_o - \bm{k}_i$. One can therefore assign to each pixel values for $q_z = \bm{q} \cdot \hat{z}$ and $q_r = \left | \bm{q} - q_z \hat{z} \right |$.

The raw scattering image is then re-gridded to an appropriate region of the reciprocal space, $(q_r,q_z)$. For each pixel in the target space, the algorithm identifies the corresponding point in the starting image, and the closest 9 pixels are averaged to yield the final intensity value. It is easily shown that the accessible range in reciprocal space is triangle-shaped, with an angle $2\alpha$ at the origin. Finally, the resulting image is divided by the form factor of the particle to yield the structure factor.

As an illustration, we show in Figure~\ref{fig:redr} re-gridded data at four different incidence angles $\alpha$ for the sample with $n = 1.976 10^{-3} \un{\AA}^{-2}$. A modulation along $q_z$ is clearly visible, indicating the presence of the $S_1$ component and hence of an interaction between layers. The images also display a linear slope in $q_z$, probably due to imperfect background subtraction. This artefact is removed prior to further treatment.

\begin{figure}
\centerline{\includegraphics[width=9cm,angle=0]{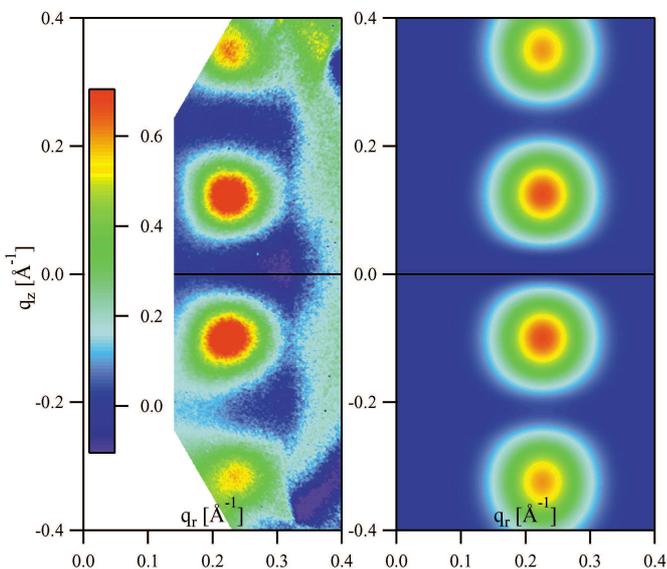}}
\caption{Comparison between the experimental data for $S(q_r,q_z) - S(q_r,q_z=0)$ derived from the image taken at $\alpha = 60{\,}^{\circ}$ in Figure~\ref{fig:redr} and the model (\ref{eq:develop_final}).} \label{fig:comp73}
\end{figure}

Although the physically relevant functions are $S_0 (q_r)$ and $S_1
(q_r)$, the most easily accessible quantity is the ``equatorial''
cut through reciprocal space $S_{\text{eq}} = S(q_r,q_z=0) = S_0 + 2
S_1$ obtained under normal incidence ($\alpha = 0$ in Figure
\ref{fig:redr}). We therefore subtract $S_{\text{eq}}$ from the
re-gridded data (an example is shown in Figure~\ref{fig:comp73},
left) and fit it with an appropriately modified version of Equation
(\ref{eq:develop_final}). The best fit for the data in Figure
\ref{fig:comp73} is given in the right image.

The vertical streak on the right in the experimental data is due to
an imperfectly subtracted diffuse scattering ring due to a kapton
window in the beam (we found that proper background subtraction is
quite difficult for tilted samples, i.e. for $\alpha \neq 0$).
Otherwise, the agreement between data and the fit is good,
showing that the model is accurate (in particular, there is no need
to include higher-order partial structure factors).

\begin{figure}
\centerline{\includegraphics[width=9cm,angle=0]{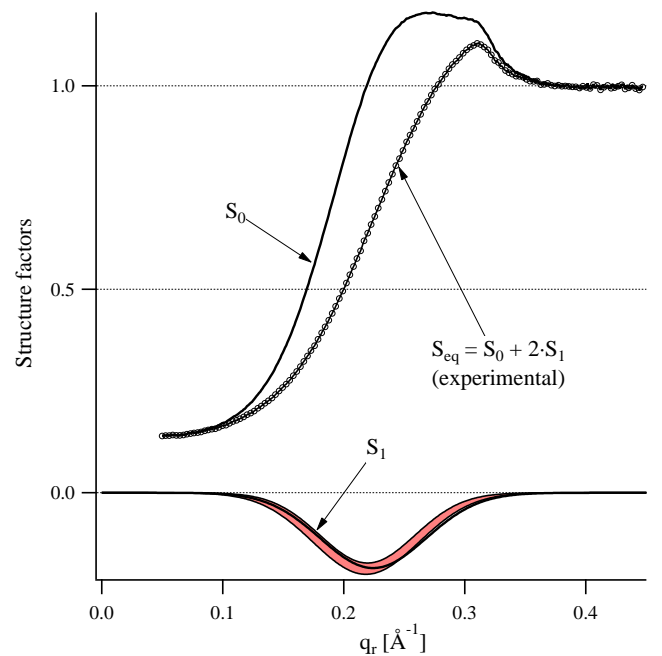}}
\caption{Partial structure factors for the sample
with $n = 1.976 10^{-3} \un{\AA}^{-2}$, extracted from the data in Figure~\ref{fig:redr}. Symbols: the equatorial structure factor $S_{\un{eq}}$ (experimental data). Lines: the structure factors $S_0$ and $S_1$. For $S_1$, the solid line represents the estimate extracted from the configuration with $\alpha = 60 {\,}^{\circ}$, while the shaded area around the curve represents the uncertainty (see text).} \label{fig:S73}
\end{figure}

This treatment yields $S_1 (q_r)$, which is then subtracted from the
equatorial structure factor $S_{\text{eq}} (q_r)$ to yield $S_0
(q_r)$. Figure~\ref{fig:S73} shows these functions for the most
concentrated sample (extracted from the data in Figure
\ref{fig:redr}). In the case of $S_1$, the solid line represents the
best fit to scattering data taken for $\alpha = 60
{\,}^{\circ}$ (Figure~\ref{fig:comp73}), while the shaded area contains the curves obtained
for $\alpha = 45$, 60 and $75 {\,}^{\circ}$, providing a rough estimate of the experimental uncertainty.

\subsection{Interaction within the layer}
\label{qr}

Once the experimental data for the partial structure factors $S_0(q_r)$ and $S_1(q_z)$ (or, equivalently, $S_{\text{eq}}$ and $S_1$) is available, it can be described in terms of the interaction potentials $V_0(r)$ and $V_1(r)$ via well-known methods in the theory of liquids, presented in Appendix~\ref{model}. The main goal of this work is determining the membrane-mediated potential $V_0(r)$. The interbilayer interaction $V_1(r)$, although needed for a complete analysis, is inessential and, furthermore, is probably very sensitive to the swelling of the phase. We therefore use a single functional form for it, given by a linear decrease from a maximum $U_1$ when the particles have the same in-plane position $\bm{r}$ to 0 when they are a distance $\xi _1$ away. After extensive tests we found that, for various shapes of the in-plane potential $V_0(r)$, the best fit for $V_1(r)$ is described by an amplitude $U_1$ around $1 \, k_B T$ and a range  $\xi _1$ of about $25 \, \text{\AA}$.

\begin{figure}
\centerline{\includegraphics[width=9cm,angle=0]{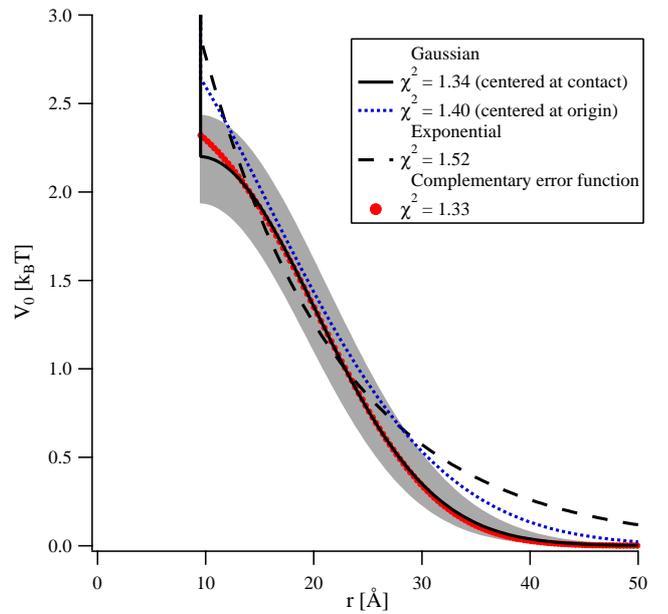}}
\caption{Best results for some of the functional forms used for $V_0(r)$ and corresponding $\chi ^2$ values. For the Gaussian centered at contact, we also plot the uncertainty range obtained as described in Appendix \ref{subsec:conf}.} \label{fig:vr0}
\end{figure}

In the following, we concentrate on the parameters describing $V_0(r)$. We considered several functional forms, such as an exponential or a linear decrease, a Gaussian centered at the origin or at the contact and a complementary error function. The best result for each of them is shown in Figure \ref{fig:vr0}, except for the linear decrease which yields much worse fits.

\begin{figure*}
\centerline{\includegraphics[width=18cm,angle=0]{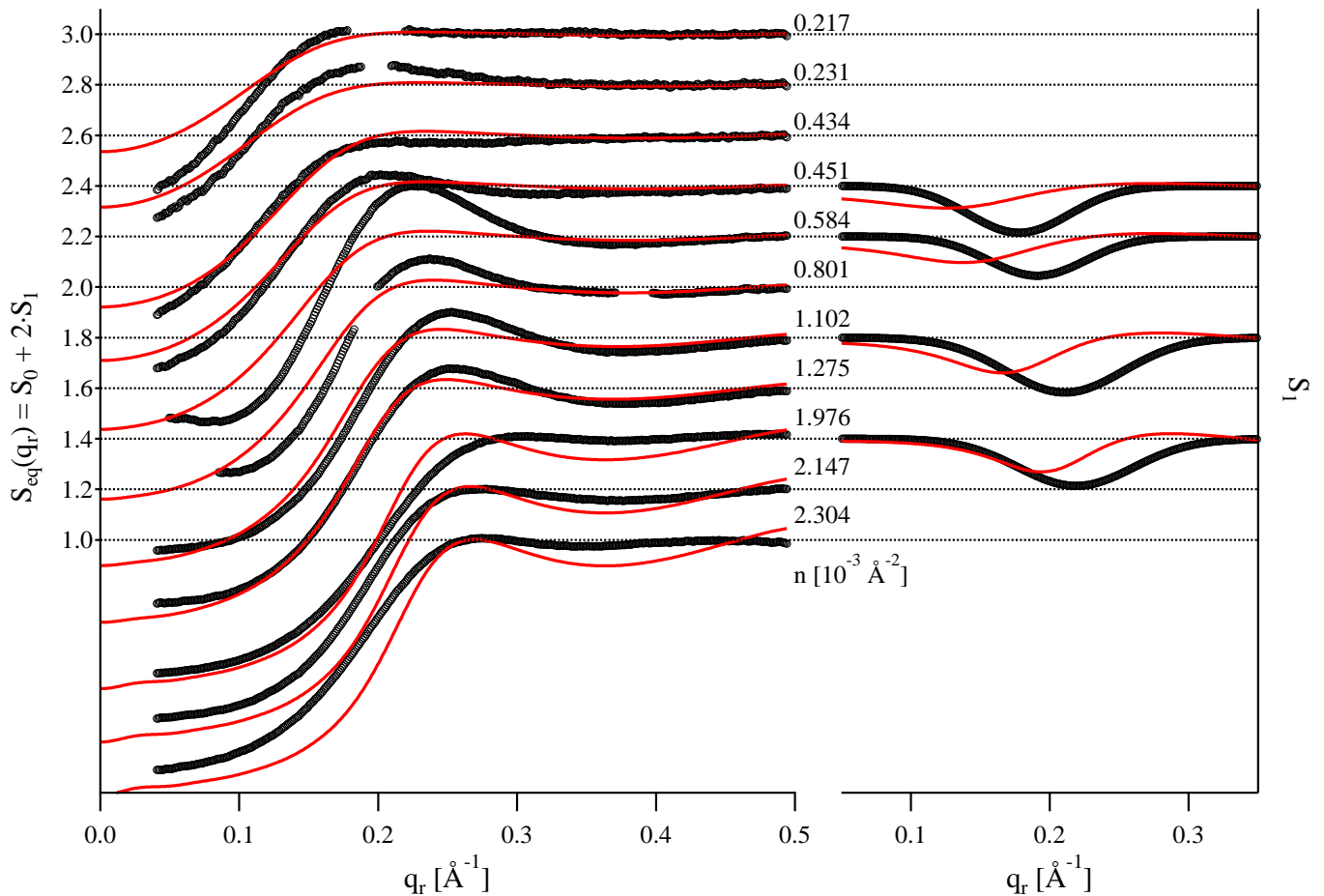}}
\caption{Best fits (lines) to the experimental data (symbols) with the model described in Section \ref{qr}. $V_0(r)$ is a Gaussian centered at contact (solid line in Figure \ref{fig:vr0}.)} \label{fig:All_graph}
\end{figure*}

We obtained very good results for the Gaussian centered at contact: $\displaystyle V_0(r) = U_c \exp \left [ - \frac{1}{2} \left ( \frac{r - 2 R}{\xi} \right ) ^2 \right ]$ (solid line in Figure \ref{fig:vr0}; the corresponding structure factors are plotted in Figure \ref{fig:All_graph}, along with the experimental data) with fit parameters $U_c = 2.2 \pm 0.2 \, \un{k_B T}$ and $\xi =10.8 \pm 0.8 \, \text{\AA}$ while the hard disk radius $R$ is fixed at 4.7~\AA. The shaded area corresponds to the fit uncertainty, estimated as in Appendix \ref{subsec:conf}.

\subsection{Hydrophobic matching}
\label{adapt}

In the treatment above, we considered that the interaction potential is independent of the particle concentration. However, one can infer from a simple elastic theory (and confirm by experimental investigations \cite{Constantin:2009}) that the amplitude of the interaction decreases with the concentration of inclusions; briefly, inserting a new particle is easier when the bilayer is already deformed by existing inclusions. Accounting for this effect requires introducing a new fit parameter, the concentration $n_0$, for which the interaction decreases significantly. We will consider the simplest model, whereby the membrane deformation is linear (and thus the elastic energy is quadratic) in the concentration: $V_0(n,r) = V_0(0,r)(1 - n/n_0)^2$ for $n < n_0$ and $V_0(n,r) = 0$ for $n \geq n_0$. We take $V_1(r)$ as independent of the concentration. As before, the in-plane potential is described by a Gaussian centered at contact. The fit is significantly improved, yielding $\chi ^2 = 1.00$, for parameters $U_c = 5.0 \pm 0.6 \, \un{k_B T}$, $\xi =9.3 \pm 0.7 \, \text{\AA}$ and $n_0 \simeq 5 \, 10^{-3} \text{\AA}^{-2}$ (roughly twice the highest experimental value available).

We emphasize that the values presented above correspond to the highly diluted limit $V_0(0,r)$, i.e.\ to the interaction between two particles in the absence of any other; this is the relevant case for comparison with most theoretical and numerical models.

One can see this decrease of the interaction potential between two particles due to the concentration (i.e.\ to the presence of other particles) as a crude way of accounting for many-body effects. However, this approximation is not controlled and its accuracy can only be verified by more complete theoretical models or by numerical simulations \cite{Harroun:1999b}.

\section{Discussion and conclusion}
\label{disc}

\begin{figure}
\centerline{\includegraphics[width=9cm,angle=0]{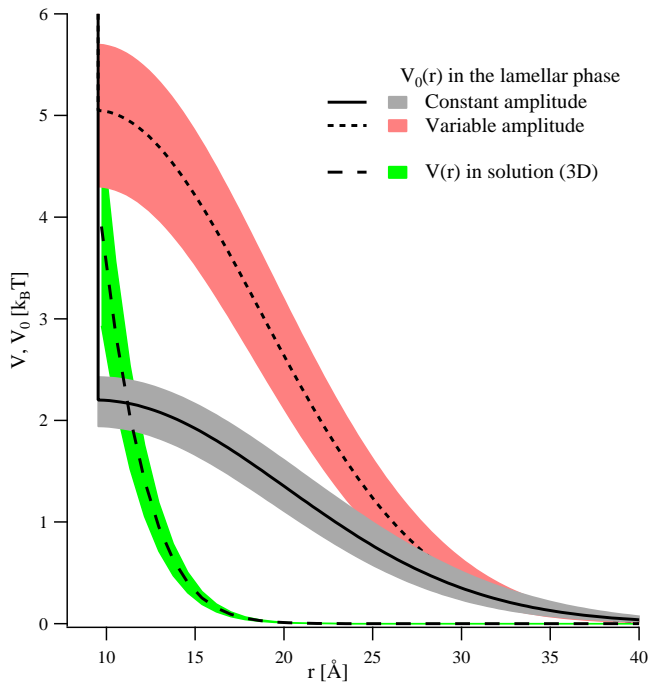}}
\caption{Best results for the in-plane interaction potential $V_0(r)$ in the lamellar phase, with constant amplitude (solid line -- see Section \ref{qr}) and with variable amplitude (dotted line -- see Section \ref{adapt}) as well as for the interaction potential $V(r)$ in solution (dashed line -- see Section \ref{solution}.) The shaded areas represent the uncertainty, determined as explained in Appendix \ref{subsec:conf}.} \label{fig:comp}
\end{figure}

The main results of this work are the in-plane interaction potential of the BuSn12 particles, $V_0(r)$ in the lamellar phase and $V(r)$ in solution; they are summarized in Figure \ref{fig:comp}. One can see straightaway that the uncertainty is small compared to the amplitude of the curves and to the difference between them, except very close to contact.

The qualitative conclusions to be drawn are similar to those already obtained in Reference \onlinecite{Constantin:2008} using a simplified model, although the amplitudes are somewhat different. The interaction potential in solution is short-ranged (about 5~{\AA} from contact), but at short distances it becomes stronger than the in-plane repulsion estimated with the constant-amplitude model ($4 \, k_BT$, as opposed to $2.2 \, k_BT$ at contact). Notwithstanding the uncertainty and the non-additivity of the potentials, it is tempting to speculate on a possible non-monotonic dependence of the ``effective'' membrane-mediated interaction, defined by subtracting the interaction in solution from that in the lamellar phase.

\appendix

\section{Reproducibility}
\label{reprod}

\begin{figure}
\centerline{\includegraphics[width=9cm,angle=0]{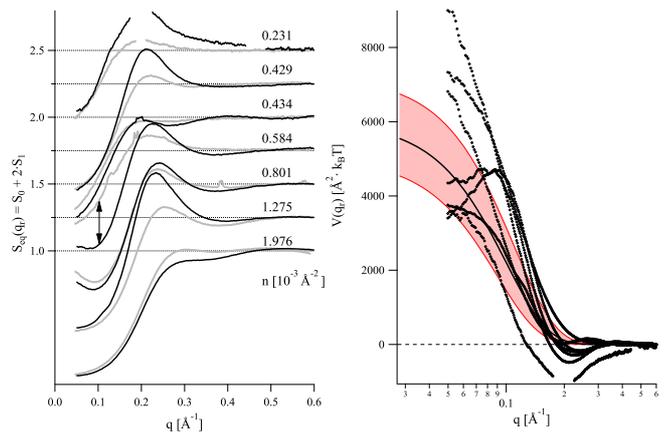}}
\caption{Left: comparison between the equatorial structure factor $S_{\text{eq}} (q_r)$ determined in two different experimental runs (black and gray curves). A noticeable difference appears for $n=0.584$, signaled by the double-headed arrow. Right: comparison between the
corresponding potentials $V(q_r)$, estimated via the RPA
approximation.} \label{fig:reprod}
\end{figure}

Since the signals to be measured are fairly weak and unstructured, the question of their reproducibility is of the utmost importance. We therefore measured a few samples twice (about one year apart), in the same experimental conditions (see Section~\ref{matmeth}), and in normal incidence, so that only the equatorial structure factor $S_{\text{eq}} (q_r)$ is available. The results are shown in Figure~\ref{fig:reprod} (left). Gray lines: earlier data, used in the analysis of Ref.~\onlinecite{Constantin:2008}. Black lines: more recent data, measured on the same samples. For clarity, the curves are shifted vertically in steps of 0.25.

Notwithstanding the discrepancy close to the peak position, the agreement is quite good, especially at small angles where $S$ departs significantly from 1. Indeed, this is the range where the interaction potential plays a significant role. For coherence, the comparison is done using the same simplified treatment as in Ref.~\onlinecite{Constantin:2008}, i.~e.\ neglecting the interaction between layers $V_1(r)$ and obtaining the Fourier transform $V(q_r)$ of the remaining interaction potential $V_0(r) \equiv V(r)$ via the RPA approximation; see Ref.~\onlinecite{Constantin:2008} for the details.

The two sets of data for $V(q_r)$ are shown in Figure~\ref{fig:reprod} (right). The solid line and the shaded area correspond to the results of Ref.~\onlinecite{Constantin:2008} (average value and uncertainty) while the symbols are obtained directly from the more recent values of $S_{\text{eq}} (q_r)$ (black lines in the left panel). Although the new data seems to yield a stronger interaction than the old, the difference is within the estimated uncertainty range. We conclude that the measurement method is reliable and that the samples did not age significantly over a period of one year.

\section{Interaction model}
\label{model}

\subsection{Analytical development}

The interaction is treated in the framework of the Ornstein-Zernike relation, with the Percus-Yevick closure approximation. The structure factors $S_{m}(q)$ are computed using the method introduced by Lado \cite{Lado:1967,Lado:1968}. An implementation for the case of no interbilayer interaction has already been used \cite{Constantin:2009}. In the present case, we extend the model to the case of several interacting bilayers. As discussed in the main text, the system can be mapped to a multi-component fluid, where each component corresponds to particles inserted within one bilayer and all particles formally occupy the same two-dimensional space. Throughout this analysis, only the two-body interaction is taken into account.

This system has been studied for a long time \cite{Lebowitz64,Baxter70}. In the following, we will use the notations of Ref.~\onlinecite{Baxter70}, except for the interaction potential, denoted here by $V$ (instead of $\phi$) for coherence with previous work. We consider $n$ bilayers (indexed by lowercase Greek indices) with periodic boundary conditions (bilayer $n$ is identical to bilayer 0). The relevant equations are:

The definition of the \textit{total correlation function} $h(r)$ in terms of the radial distribution function $g(r)$:
\beq
h_{\alpha \beta} (r) = g_{\alpha \beta} (r) - 1
\label{eq:hr}
\eeq

The Ornstein Zernike relation, defining the \textit{direct correlation function} $c(r)$:
\beq
h_{\alpha \beta} (r) = c_{\alpha \beta} (r) + \sum_{\gamma} \rho_{\gamma} \int {\dd}^2 \bm{s} \, c_{\alpha \gamma} (|\bm{s}|) h_{\gamma \beta} (|\bm{r - s}|)
\label{eq:OZ}
\eeq
\noindent where $\rho_{\gamma} = \rho$ is the number density in the plane of the bilayers and $\gamma$ runs over all bilayers. Alternatively, the equation above can be rewritten in reciprocal space, yielding:
\beq
h_{\alpha \beta} (q) = c_{\alpha \beta} (q) + \sum_{\gamma} \rho_{\gamma} c_{\alpha \gamma} (q) h_{\gamma \beta} (q)
\label{eq:OZq}
\eeq

The Percus-Yevick closure:
\beq
c_{\alpha \beta} (r) = [1+h_{\alpha \beta} (r)] \left [ 1 - \exp \left ( \frac{V_{\alpha \beta} (r)}{k_BT} \right ) \right ]
\label{eq:PY}
\eeq
\noindent with $V_{\alpha \beta} (r)$ representing the interaction potential between particles situated in bilayers $\alpha$ and $\beta$. For simplicity's sake, we consider that the particles only interact within the same bilayer and with particles in the nearest-neighbor bilayers ($V_{\alpha \beta} (r) \equiv 0$ for $|\alpha - \beta|>1$). Furthermore, translation symmetry (an identical environment for all bilayers, i.e.\ periodic boundary conditions) and mirror symmetry ($V_{\alpha \beta} \equiv V_{\beta \alpha}$, and similar relations for functions $c$ and $h$) allow us to define all functions with respect to layer 0, i.e. $\alpha = 0$ throughout.

The equations above must be solved numerically; for convergence reasons, it is much better to work with the \textit{indirect correlation function} (denoted in Ref.~\onlinecite{Lado:1967} by $H(r)$), $\gamma _{0\beta}(r) = h_{0\beta}(r) - c_{0\beta}(r)$ instead of the total correlation function.

As an example, for $n=3$ bilayers (with periodic boundary conditions, $3 \equiv 0$), equations (\ref{eq:OZq}) and (\ref{eq:PY}) can be written out explicitly as:

\begin{subequations}\label{eq:alleq}
\begin{eqnarray}
\gamma _{00}(q) &=& \rho \left [ c_{00} (\gamma _{00} + c_{00}) + 2 c_{01}(\gamma _{01} + c_{01})\right ]\\
\gamma _{01}(q) &=& \rho \left [ c_{00} (\gamma _{01} + c_{01}) + c_{01} (\gamma _{00} + c_{00}) + c_{01} (\gamma _{01} + c_{01}) \right ]\nonumber\\
&&\\
c_{00}(r) &=& [1 + \gamma _{00}] [\exp (-V_{00}/k_BT) - 1]\\
c_{01}(r) &=& [1 + \gamma _{01}] [\exp (-V_{01}/k_BT) - 1]
\end{eqnarray}
\end{subequations}
\noindent where we omitted the argument of the functions on the right-hand side. One starts by applying in real space equations (\ref{eq:alleq}c-d) with reasonable initial guesses, and then solving (\ref{eq:alleq}a-b) in reciprocal space. The procedure is iterated until stability is reached. The experimentally relevant structure factors are simply: $S_0(q)=1+ h_{00}(q)$ and $S_1(q)=h_{01}(q)$. For completeness, the solution of (\ref{eq:alleq}a-b) is:

\begin{subequations}\label{eq:gamma}
\begin{eqnarray}
\gamma _{00}(q) &=& \frac{-c_{00}^2 (c_{00} + c_{01} - 1/\rho) + 2 c_{01}^2(c_{00} + 1/\rho)}
{(c_{00} - 1/\rho)^2 + c_{01} (c_{00} - 1/\rho) - 2 c_{01}^2}\nonumber\\
&& \\
\gamma _{01}(q) &=& c_{01} \frac{-(c_{00} + c_{01})(c_{00} - 1/\rho)+ c_{00}/\rho + 2 c_{01}^2}{(c_{00} - 1/\rho)^2 + c_{01} (c_{00} - 1/\rho) - 2 c_{01}^2}\nonumber\\
&&
\end{eqnarray}
\end{subequations}

We checked that, for relevant values of the interaction parameters, the structure factors obtained with $n=5$ and $n=7$ are almost superposed, and slightly different from those obtained for $n=3$ (with periodic boundary conditions in all cases). All results presented above are obtained with $n=7$. We also checked that the higher order structure factors are negligible, viz.\ $S_3(q) \ll S_2(q) \ll S_1(q)$, in agreement with the experimental observations.

\subsection{Software implementation}

The routine discussed above is implemented in \textsc{Igor Pro 6} as an "all-at-once" function, i.e.\ all points of the output vector are returned at the same time. This approach is particularly useful for such iterative procedures, where, in order to get its value at one point, the entire function needs to be calculated anyway. The Fourier transform was implemented as a matrix operation with pre-computed coefficients over a fixed equidistant grid. Wherever possible, the wave operations were done using the \texttt{MatrixOp} command.

For each combination of fit parameters, the structure factors $S_0$ and $S_1$ are calculated for all densities. Comparison with the experimental data yields the goodness-of-fit function $\chi ^2$, which is minimized using the \texttt{Optimize} operation with the simulated annealing method.

\subsection{Confidence range}
\label{subsec:conf}

Once a minimum is found, the goodness-of-fit function $\chi ^2$ is plotted as a function of the parameters $U_0$ (or $U_c$) and $\xi$.

For the 3D case (interaction in solution), one notices that the minimum is in fact an extended valley, covering a wide range of parameters, and roughly defined as the locus of the points where the integral of the potential is constant. This is understandable, since the fit is mainly sensitive to the low values of the scattering wave vector $q$, where it is affected by $V(q \rightarrow 0)$, i.e.\ by the integral of $V(r)$, rather than by its finer details. We therefore choose a number of points along this valley, with close to minimum $\chi ^2$, and trace the corresponding $V(r)$ instances. The confidence range (shown as a shaded area in Figure \ref{fig:V_3D}, bottom) is chosen by manually adjusting the parameters to yield lower and higher envelopes to this sheaf of curves.

In the 2D case (interaction within the layer), the minimum is well localized in the $(U_c, \xi)$ plane. However, a standard statistical estimation of the confidence range is irrelevant, since it would yield extremely tight confidence ranges. Indeed, the discrepancy between model and data is mainly due to systematic errors and (presumably) to an inexact functional form for the interaction potential. In order to account for the latter effect we took as an acceptable increase $\Delta \chi ^2$ the difference between the Gaussian centered at contact ($\chi ^2 = 1.34$) and that centered at the origin ($\chi ^2 = 1.40$), both plotted in Figure \ref{fig:vr0}. The uncertainty on $U_c$ and $\xi$ given in Section \ref{qr} is determined based on this numerical value. For comparison, we also show the exponential decay with $\chi ^2 = 1.52$. We follow a similar procedure for the variable amplitude model discussed in Section \ref{adapt}, with $\Delta \chi ^2 = 0.1$.

\bigskip

\begin{acknowledgments}
The ESRF is gratefully acknowledged for the award of beam time (experiment 02-01-756) and we thank C. Rochas for competent and enthusiastic support. F. Ribot is acknowledged for providing the BuSn12 particles and for edifying discussions.
\end{acknowledgments}

\newpage

%

\end{document}